\documentclass[smallextended]{svjour3}     
\smartqed  
\usepackage[nice]{nicefrac}
\usepackage{graphicx}
\usepackage{wrapfig}
\usepackage{url}
\newcommand\eat[1]{}
\usepackage{tikz}
\usepackage{booktabs}  
\usetikzlibrary{arrows}
\journalname{}

\usepackage{array,xspace,multirow,hhline,graphicx,xcolor,tikz,colortbl,tabularx,amsmath,amssymb,amsfonts}
\usepackage[round]{natbib}
\usetikzlibrary{shapes}
\usepackage{algorithm,algorithmic}
\usepackage{mathrsfs}
\usepackage{enumitem}

\usepackage{pifont}
\setenumerate[1]{label=(\emph{\roman{*}}),ref=(\emph{\roman{*}}),leftmargin=*}

\usepackage{txfonts}
\usepackage{eucal} 

\usepackage{tikz}

\newlength{\wordlength}

	\newcommand{\floor}[1]{\lfloor #1 \rfloor }

\newcommand{\Pref}[1][]{
	\ifthenelse{\equal{#1}{}}{\mathrel \succsim}{\mathop{\succsim_{#1}}}
}                                          
\newcommand{\sPref}[1][]{                  
	\ifthenelse{\equal{#1}{}}{\mathrel \succ}{\mathop{\succ_{#1}}}
}                                          
\newcommand{\Indiff}[1][]{                 
	\ifthenelse{\equal{#1}{}}{\mathrel \sim}{\mathop{\sim_{#1}}}
}
\newcommand{\prefset}[1][]{\ifthenelse{\equal{#1}{}}{\mathcal{\succsim}}{\mathcal{\succsim}_{#1}}}

\newcommand{\pav}[0]{\ensuremath{\mathit{PAV}}\xspace}

\newcommand{\rav}[0]{\ensuremath{\mathit{SeqPAV}}\xspace}
\newcommand{\gspav}[0]{\ensuremath{\mathit{GroupSeqPAV}}\xspace}

\newcommand{\jr}{\ensuremath{\mathit{JR}}\xspace}
\newcommand{\ejr}{\ensuremath{\mathit{EJR}}\xspace}

\newcommand{\pjr}{\ensuremath{\mathit{PJR}}\xspace}
\newcommand{\cjr}{\ensuremath{\mathit{CJR}}\xspace}

\newcommand{\calA}{{\vec{A}}}

\let\enumtemp=\enumerate
\def\enumerate{\enumtemp\itemsep 1pt}
\let\itemtemp=\itemize
\def\itemize{\itemtemp\itemsep 1pt}

\newcommand{\Omit}[1]{}

\begin{document}

\title{Proportional Representation in \\Approval-based Committee Voting and Beyond\thanks{}}


	\author{Haris Aziz}


	\institute{%
	  Haris Aziz \at
	  Data61, CSIRO and UNSW
	Sydney, Australia \\
	  Tel.: +61-2-9490\,59090 \\
	  Fax: +61-2-8306\,0405 \\
	  \email{haris.aziz@data61.csiro.au}	  
	         }


\maketitle

\begin{abstract}
Proportional representation (PR) is one of the central principles in voting. 
Elegant rules with compelling PR axiomatic properties have the potential to be adopted for several important collective decision making settings.
I survey some recent ideas and results on axioms and rules for proportional representation in committee voting. 
\end{abstract}
	
	\keywords{Proportional Representation, Justified Representation, Multi-winner Voting, Committee Voting}

\noindent
\textbf{JEL Classification}: C70 $\cdot$ D61 $\cdot$ D63 $\cdot$ D71

\section{Introduction}


When making collective decisions, fairness entails that the decision is made in accordance with the will and desire of the people and that each person has equal influence. A natural principle that captures this requirement is proportional representation: the bigger a group, the more representation it should have. This general principle of proportionality is engrained in just societies.\footnote{Aristotle said ``[...] \emph{what the just is-the proportional; the unjust is what violates the proportion}.'' (Nicomachean Ethics, Written 350 B.C).}

We discuss the issue of proportional representation in the context of approval-based committee voting (also called multi-winner voting with approvals).  The setting involves a set $N=\{1,\ldots, n\}$ of voters and a set $C$ of candidates. Each voter $i\in N$ submits an approval ballot $A_i\subseteq C$, which represents the subset of candidates that she
approves. We refer to the list $\calA = (A_1,\ldots, A_n)$ of approval ballots as the {\em ballot profile}. Based on the approval of the voters, the goal is to select  a target $k$ number of candidates. 

The setting has inspired a number of natural voting rules (see e.g. the survey by \citet{Kilg10a}).
Many of the voting rules are designed with the goal of achieving some form of just representation. However it is not entirely obvious what axiom captures proportional representation requirements.

How should proportional representation be defined in approval-based committee voting?
We first note that it can be defined in a straightforward manner for a restricted version of approval-based committee voting that we will refer to as `\emph{polarized}'. In a polarised profile, voters can be partitioned into disjoint groups such that the approvals of voters in the same group coincide, and approvals from two different groups do not intersect. For polarized preferences, the proportional representation requirement can easily be formalized as follows: for any group $G$ that approves candidates in set $C_G$, we can require that at least $\min({\floor{k\frac{|G|}{n}}},|C_G|)$ candidates from $C_G$ are selected. Not only can the requirement be easily defined, it can also achieved by the following rule:
\begin{quote}
\emph{\textbf{\gspav}: Sequentially select candidates to be placed in the committee. In each round, consider the group $G$ that has the largest value $|G|/(r(G)+1)$ (where $r(G)$ is the current number of representatives of $G$) and still has an approved candidate $c$ that is yet not selected. Place $c$ in the committee. Repeat until $k$ candidates are selected. }
\end{quote}

Polarized preferences are typically prevalent in `closed list' party elections in which voters vote for parties and each party gets seats in proportion of votes~\citep{Jans16a}. 
These seats are then filled up by representatives from the corresponding party. If each party has sufficient number of representatives, then the problem reduces to giving each party at least the integer part of the target quota and then \emph{apportion} the remaining seats~\citep[see e.g., ][]{BLS16a,SFF16a}. There are several ways to do this and there is a substantial body of work on proportional representation via apportionment~\citep[see e.g., ][]{BaYo01a,Puke14a,PeTe90a}. 
In this restricted setting that models `closed list' party elections,  \gspav corresponds to the D’Hondt method (also called the Jefferson method) for apportionment.

Representation becomes more challenging to formalize when voters in a group may approve candidates approved by voters outside the group. 
The challenge stems from the fact that 
the approval-based committee voting setting does not even assume pre-specified groups since each individual voter is free to approve any subset of candidates. In what follows we describe recent work on formalising proportional representation axioms that are referred to as justified representation axioms.

\section{Justified Representation Properties}

We present justified representation axioms that are all based on the proportionality representation principle. The idea behind all the axioms is that a cohesive and large enough group of voters deserves sufficient number of approved candidates in the winning set of candidates. 


%


\begin{definition}[Justified representation (\jr)]
Given a ballot profile $\calA = (A_1, \dots, A_n)$ over a candidate set $C$ and a target committee size $k$,
we say that a set of candidates $W$ of size $|W|=k$ {\em satisfies justified representation 
for $(\calA, k)$} if 
$\forall X\subseteq N: |X|\geq \frac{n}{k} \text{ and } |\cap_{i\in X}A_i|\geq 1 \implies (|W\cap (\cup_{i\in X}A_i)|\geq 1).$
\jr was proposed by \citet{ABC+15a,ABC+16a}. 
\end{definition}

The rationale behind \jr is that if $k$ candidates are to be selected, then, intuitively,
each group of $\frac{n}{k}$ voters ``deserves'' a representative. Therefore, a set of $\frac{n}{k}$ voters 
that have at least one candidate in common should not be completely unrepresented. \jr can be strengthened to \pjr and \ejr.

\begin{definition}[Proportional Justified Representation (\pjr)]
Given a ballot profile $(A_1, \dots, A_n)$ over a candidate set $C$, a target committee size $k$, $k\le m$, and integer $\ell$
we say that a set of candidates $W$, $|W|=k$, {\em satisfies $\ell$-proportional justified representation
for $(\calA, k)$}  if
$\forall X\subseteq N: |X|\geq \ell\frac{n}{k} \text{ and } |\cap_{i\in X}A_i|\geq \ell \implies (|W\cap (\cup_{i\in X}A_i)|\geq \ell).$

We say that $W$ {\em satisfies proportional justified representation for $(\calA, k)$} if it {satisfies $\ell$-proportional justified representation
for $(\calA, k)$} and all integers $\ell\leq k$. \pjr was formally studied by \citet{SFF+17a}. 
\end{definition}

\begin{definition}[Extended justified representation (\ejr)]
Given a ballot profile $(A_1, \dots, A_n)$ over a candidate set $C$, a target committee size $k$, $k\le m$,
we say that a set of candidates $W$, $|W|=k$, {\em satisfies  $\ell$-extended justified representation
for $(\calA, k)$}  and integer $\ell$ if
$\forall X\subseteq N: |X|\geq \ell\frac{n}{k} \text{ and } |\cap_{i\in X}A_i|\geq \ell \implies (\exists i\in X: |W\cap A_i|\geq \ell).$

We say that $W$ {\em satisfies extended justified representation for $(\calA, k)$} if it {satisfies $\ell$-extended justified representation
for $(\calA, k)$} and all integers $\ell\leq k$.
\ejr was proposed by \citet{ABC+16a}.
\end{definition}


%
%


It is easy to observe the following relations: $\ejr \implies \pjr \implies \jr.$
 Also note that if we only consider $\ell=1$ in the definitions of \pjr, and \ejr we get \jr. We also observe that for $k=1$, \jr, \pjr, and \ejr are equivalent. 
 
 %
 
 \section{Achieving Proportional Representation}
 We say that a rule satisfies \jr/\pjr/\ejr if it always returns a committee satisfying the corresponding property. For preferences that are not polarized, the definition of \gspav needs to be extended since there are no clear-cut groups for general approval ballots. One such generalisation is 
 called \rav. Let $H$ be a function defined on integers such that $H(p)=0$ for $p=0$ and  $H(p)=\sum_{j=1}^p\frac{1}{j}$ otherwise. Let the \pav score of a committee $W$ be $ \sum_{i\in N}H(|W\cap A_i|)$. Then the \rav rule is defined as follows.

%
%

\begin{quote}
	\textbf{\rav}: Set $W=\emptyset$.
	Then in round $j$, $j=1, \dots, k$, add a new candidate to $W$ so that
	the \pav score of $W$ is maximised. 
	%
	\end{quote} 
	
	\rav was originally proposed by \citet{Thie95a}. 
Although \rav seems like a reasonable extension of \gspav, it has been shown that \rav does not even satisfy \jr~\citep{ABC+16a}. Incidentally, \rav is not the only rule that may violate \jr. \citet{ABC+16a} pointed out that several well-known rules that are designed for representation purposes fail to satisfy \jr.\footnote{There is a natural dual version of \rav called \emph{RevSeqPAV} (in which candidates are iteratively deleted from $C$ that leads to minimal decrease in total \pav score) which also violates \jr.}

	Whereas SeqPAV iteratively builds a committee while trying to maximize 
	the \pav score, one could also try to find a committee that globally maximizes the \pav score. Such a rule is popularly known as \pav and was originally proposed by \citet{Thie95a}.
	%
In contrast to \rav, \pav always returns a committee that satisfies \ejr~\citep{ABC+16a} thereby giving a constructive argument for the existence of a committee that satisfies \ejr. 
Although \pav satisfies \ejr, it does have some drawbacks. 
From a computational perspective, finding a \pav outcome is NP-hard~\cite{AGG+15a,SFL16a}. The computational intractability renders the rule impractical for large scale voting.
From an axiomatic perspective, \pav does not satisfy certain desirable axioms such as committee monotonicity.\footnote{Committee monotonicity requires that for any outcome $W$ of size $k$, there is a possible outcome $W'$ of size $k+1$ such that $W'\supset W$.}


When \ejr was proposed it was not clear whether it can be achieved in polynomial time. In view of this, researchers turned to designing polynomial-time algorithms to achieve the weaker property of \pjr. \citet{BFJL16a} proved that SeqPhragm\'{e}n (an algorithm proposed by Swedish mathematician Phragm\'{e}n in the 19th century) is polynomial-time and returns a committee satisfying \pjr.  Independently and around the same time as the result by \citet{BFJL16a}, \citet{SFF16a} presented a different algorithm that finds a \pjr committee and also satisfies other desirable monotonicity axioms. Like \rav, both algorithms sequentially build a committee while optimising a corresponding load balancing objective. However the algorithms may not return a committee that satisfies \ejr. 

Recently, three different groups~\citep{AzHu17a,SLES17a,SEL17a} have independently and around the same time shown that a committee satisfying \ejr can be computed in polynomial time.\footnote{Although a committee satisfying \ejr can be computed in polynomial time, testing whether a given committee satisfies a representation property is coNP-complete for both \ejr~\citep{AGG+15a,ABC+16a} and \pjr~\citep{AzHu16a}.} Two of the groups~\citep{AzHu17a,SLES17a} have essentially the same idea of maximizing the \pav score via local search and implementing swaps of candidates. 

\section{Discussion}

We focussed on proportional representation under approvals and discussed natural axioms for this purpose. 
It will be interesting to see how ideas from recent developments can be used to design voting rules that are compelling for proportional representation for dichotomous preferences as well as more general preferences. For example, it will be interesting to design or identify rules that satisfy a strong notion of proportional representation along with other natural axioms such as candidate monotonicity\footnote{Candidate monotonicity requires that increasing the support for candidate should never make a selected candidate unselected.} and committee monotonicity.

When considering approvals, \ejr can be further strengthened to \cjr (core justified representation).  Given a ballot profile $(A_1, \dots, A_n)$ over a candidate set $C$, a target committee size $k$, $k\le m$,
we say that a set of candidates $W$, $|W|=k$, satisfies \emph{core representation (\cjr)} if there exists no coalition $X\subseteq N$ such that $|X|\geq \ell n/k$ and there is a set $D\subset C$ such that $|D|=\ell$ and $|A_i\cap D| > |A_i\cap W|$ for each $i\in X$. 
We call such a coalition $X$ as a \cjr blocking coalition.
A core concept equivalent to \cjr but formalized in a different way was discussed by  \citet{ABC+16a}. 
It is interesting that core stability, one of the central ideas of economic design is also meaningful in the context of proportional representation.
It remains open whether a committee satisfying \cjr always exists and whether such a committee can be computed in polynomial time.\footnote{If the definition is strengthened to \emph{strict} core (a set of candidates $W$, $|W|=k$, satisfies \emph{strict core representation (\cjr)} if there exists no coalition $X\subseteq N$ such that $|X|\geq \ell n/k$ and there is a set $D\subset C$ such that $|D|=\ell$ and $|A_i\cap D| \geq |A_i\cap W|$ for each $i\in X$ and $|A_i\cap D| >|A_i\cap W|$ for some $i\in X$), one can obtain simple examples for which no stable outcome exists. } 

Considering that proportional representation for approvals (that capture dichotomous preferences) is a non-trivial task, it leads to the question of how it should be defined in the context of preferences that are not dichotomous. 
The axioms \jr, \pjr, and \ejr can also be extended to the case where voters have strict or weak orders over candidates. However for a natural generalisation of \jr to the case of linear orders, it turns out that not only a committee satisfying the property may not exist, it is also NP-hard to compute~\citep{AEF+17a}. It will be interesting to see if compelling proportional representation axioms can be proposed for general preferences that guide the design and analysis of rules.



There is scope for substantial and fruitful research in formalizing and achieving proportional representation for more general or complex voting settings in which simultaneous or sequential decisions are made. Finally, multi-winner voting deserves a thorough research investigation with respect to goals other than proportional representational as well~\citep{FSST17a}.

\begin{acknowledgements}
	This paper was written as a companion paper to the authors's talk at the Dagstuhl Seminar on
Voting: Beyond Simple Majorities and Single-Winner Elections (25--30, June 2017).
	 The author is supported by a Julius Career Award. 
	 He thanks all of his collaborators on this topic for several insightful discussions. 
	 He also thanks Barton Lee for feedback. 
\end{acknowledgements}

%

\end{document}